\documentclass[conference]{IEEEtran}

\usepackage{cite}
\usepackage{amsmath,amssymb,amsfonts}
\usepackage{algorithmic}
\usepackage{graphicx}
\usepackage{textcomp}
\usepackage{caption}
\usepackage{multirow}
\usepackage{booktabs}
\usepackage[colorlinks=true,citecolor=blue]{hyperref}

\usepackage[dvipsnames]{xcolor}
\usepackage{colortbl}
\definecolor{lightblue}{HTML}{E0ECFF}

\def\BibTeX{{\rm B\kern-.05em{\sc i\kern-.025em b}\kern-.08em
    T\kern-.1667em\lower.7ex\hbox{E}\kern-.125emX}}
\begin{document}

\title{Towards a Single ASR Model That Generalizes to Disordered Speech}

\author{\IEEEauthorblockN{Jimmy Tobin}
\IEEEauthorblockA{\textit{Google Research} \\
Mountain View, USA \\
0009-0009-5733-9295}
\and
\IEEEauthorblockN{Katrin Tomanek}
\IEEEauthorblockA{\textit{Google Research} \\
Mountain View, USA \\
0000-0002-9873-7144}
\and
\IEEEauthorblockN{Subhashini Venugopalan}
\IEEEauthorblockA{\textit{Google Research} \\
Mountain View, USA \\
0000-0003-3729-8456}
}

\maketitle

\begin{abstract}
    This study investigates the impact of integrating a dataset of disordered speech recordings ($\sim$1,000 hours) into the fine-tuning of a near state-of-the-art ASR baseline system. %
    Contrary to what one might expect, despite the data being less than $1\%$ of the training data of the ASR system, we find a considerable improvement in disordered speech recognition accuracy.
    Specifically, we observe a $33\%$ improvement on %
    prompted speech, and a $26\%$ improvement on a newly gathered spontaneous, conversational %
    dataset of disordered speech. %
    Importantly, there is no significant performance decline on standard speech recognition benchmarks. %
    Further, %
    we observe that the proposed tuning strategy helps close the gap between the baseline system and personalized models by $64\%$ highlighting the significant progress as well as the room for improvement.
    Given the substantial benefits of our findings, this experiment suggests
    that from a fairness perspective, incorporating a small fraction of high
    quality disordered speech data in a training recipe is an easy step that
    could be done to make speech technology more accessible for users with speech disabilities.

\end{abstract}


\section{Introduction}
Automatic speech recognition (ASR) has become a %
ubiquitous technology in critical applications such as voice dictation and typing, voice assistance in devices, %
and phone based customer support systems. %
Despite efforts on building inclusive ASR models e.g. by incorporating diverse speech patterns, and data from low-resource languages and historically underrepresented groups\cite{joshi2020state, yadav2022survey}, current models can result in less than optimal performance for individuals who have speech patterns that are not represented in training. 
This has been studied for atypical or disordered speech, e.g. on individuals
with Parkinson's Disease, amyotrophic lateral sclerosis (ALS), Down Syndrome,
etc.~\cite{gutz2022validity}.

One approach to mitigate this disparity in speech recognition performance is personalization~\cite{green2021,tobin2022}, where an ASR model is finetuned with an individual's own speech recordings to improve recognition for that one individual. However, curating even small amounts of data for personalization may prove hard for certain individuals e.g., those with disorders that affect literacy, dexterity, cognition or stamina. Further, %
there may be a drift in quality~\cite{tomanek2023analysis} of the model over time incurring additional costs for the participant and for model maintenance.

A model that understands individuals with disordered speech ``out of the box" without any  set up or work incumbent on the end user is %
a significant step towards making technology more usable by people with atypical or disordered speech. %
In this work we demonstrate the viability of a single production-level speech recognition model that can be trained to serve individuals with disordered speech with no regression in performance for typical speakers.

Specifically, we assess the impact of adding disordered speech data to an ASR model's training set, providing guidelines for evaluation and creation of 
a high quality %
speaker-\textit{independent} ASR (SI-ASR) model. 
We analyze the impact of the addition of bespoke training data on the performance of recognizing both disordered as well as typical speech. %
Further, we study the components that have helped narrow the gap between SI-ASR and personalized ASR models, and the scope for improvement. %
Our work provides experimental evidence to support initiatives such as the Speech Accessibility Project~\cite{SAP} %
to gather and incorporate disordered speech data not just to develop speaker-dependent personalized models but also integrate such data directly into the development of generic ASR models that stand to benefit all users, particularly those with speech disorders.

\section{Methods}

In this work, we build on top of Google's Universal Speech Model (USM)~\cite{zhang2023google}. USM is a family of large models consisting of a Conformer~\cite{gulati2020conformer} encoder and a CTC decoder that are trained with both unlabeled and labeled data in various stages of training in an effort to build a multilingual model capable of recognizing 100+ languages. %
It was trained with data from 300+ languages sourced from public datasets as well as YouTube videos. While multi-linguality was the chief aim of training USM, disordered speech was not represented in the model's training data. Our work follows USM's tuning recipe for a 2B parameter model and aims to investigate the addition of disordered speech data in the model training and study its effect on both typical and atypical speech recognition performance.

\subsection{Datasets}
\subsubsection{\textbf{Prompted Speech Dataset}}
\label{sec:siasr_dataset}

For this work, we derive our \textit{prompted} speech dataset of disordered speech from the Euphonia corpus~\cite{macdonald2021}, a large collection of speech samples from people with different types and severity of speech disorders. The corpus consists of prompted speech samples where participants were shown phrases from different domains (e.g., "home automation" or shorter conversational phrases), which they then repeat.  The dataset includes a rich speaker-level metadata annotated by Speech and Language Pathologists (SLPs), including information such as severity on a Likert scale (ranging from "typical" to "profound"), underlying etiologies leading to the respective speech disorder and more.

\textbf{Data splitting guidelines.} We create splits of this dataset specifically for evaluating and tuning \textbf{SI}-ASR systems
using the following core design principles to ensure  generalization of the results:

\begin{enumerate}
    \item[(1)] Strictly no overlap between training and test split in terms of \textit{speakers} or \textit{phrases}.
    \item[(2)] Diverse and representative set of speakers with speech impairments in the test set.
    \item[(3)] Training set with as many speakers (and utterances) as possible without violating the previous 2 principles.
\end{enumerate}

\begin{table}[]
    \centering
    \begin{tabular}{l|l|l|l}
         \toprule \midrule
         \textbf{Category} & \textbf{Value} & \textbf{Test} & \textbf{Train} \\ \midrule

         \multirow{3}{*}{Gender} & male & 58.8\% (117) & 48.0\% (583) \\
                & female & 41.2\% (82) & 31.5\% (383) \\
                & unknown & NA & 20.5\% (249) \\
         \midrule
         \multirow{6}{*}{Severity} & typical & 8\% (16) & 12.3\% (149) \\
                  & mild & 50.8\% (101) & 34.7\% (422) \\
                  & moderate & 24.6\% (49) & 18.8\% (229) \\
                  & severe & 16.6\% (33) & 12.2\% (148) \\
                  & profound & NA & 1.5\% (18) \\
                  & unknown & NA & 20.5\% (249) \\
        \midrule
        
                 & Parkinson's & 27.1\% (54) & 12.2\% (148) \\
        Etiology & ALS & 16.6\% (33) & 18.4\% (224) \\
         (top-5) & Cerebral Palsy & 14.1\% (28) & 8.6\% (105) \\
         & Down Synd. & 8.0\% (16) & 12.2\% (148) \\
                 & stutter & 6.5\% (13) & 4.0\% (50) \\
        \bottomrule
    \end{tabular}
    \caption{Diversity metrics of our \textbf{prompted speech dataset}, reporting percentage (and absolute number) of speakers.}
    \label{tab:si_asr_diversity_stats}
\end{table}

To build the splits based on these principles, we start with the test set since it needs to be of the highest quality.
We first identify speakers and phrases to be used in the test set, ensuring diversity along different axes (Table~\ref{tab:si_asr_diversity_stats}). %
Moreover, we ensure utterances in the test set have been evaluated for quality. This
includes audio quality checks as well as transcript corrections if needed (as described in~\cite{macdonald2021} and~\cite{jiang2024learnings}). 
Each speaker in the test set has exactly 20 utterances from the \textit{home automation} and \textit{caregiver} domain; %
in addition, each speaker may have up to 20 utterances from
the  \textit{conversational} domain (if their recordings are available for this category).

\begin{figure}
\vspace{-0.2cm}
    \centering
    \includegraphics[width=\linewidth]{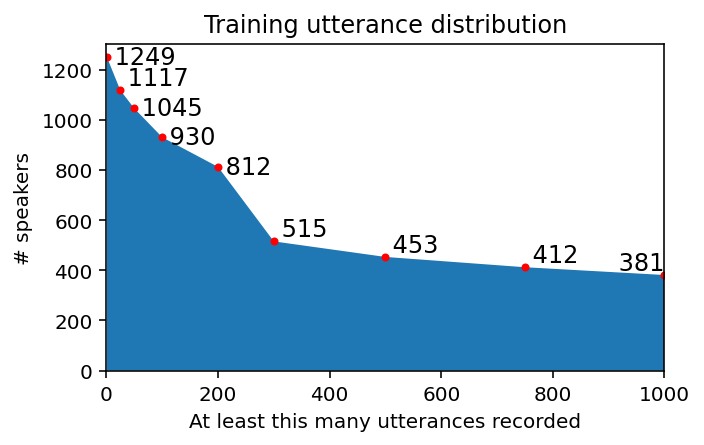}
    \caption{Distribution of number of training utterances per speaker in training split of our prompted speech dataset.}
    \label{fig:training_utterance_distribution}
\end{figure}

In the next step, we create the training set from the remainder of the Euphonia corpus, ensuring that no utterances with phrases or speakers used in the test set are added to the training split. We don't control for the number of utterances per speaker, but instead use all utterances that don't violate design principle (1).
Figure~\ref{fig:training_utterance_distribution} shows the resulting distribution of utterances per speaker in the training set.

To ensure diversity, we refer to the speaker metadata (where available) from the Euphonia corpus, including gender, etiology, and severity. Table~\ref{tab:si_asr_diversity_stats} gives an overview to the diversity metrics of the prompted speech dataset. In our dataset, gender is biased towards a much larger percentage of males. This reflects the underlying true distribution, where a larger male population have etiologies that result in disordered speech \cite{cerri2019parkinson,drayna1999sex,kovaleva2002sex,odding2006epidemiology,manjaly2010sex}. As for etiologies, only the top-5 most common etiologies are listed here. Overall, more than 20 different etiologies are covered in the test set. Splitting based on design principle (1) necessarily leads to data loss. %
Overall, our method drops around 18\% of the utterances of the underlying Euphonia corpus. Table~\ref{tab:si_asr_size_stats} shows the sizes of the resulting splits in terms of speakers and utterances.

\begin{table}[h]
\vspace{-0.1cm}
    \centering
    \begin{tabular}{l|l|l|l}
         \toprule \midrule
         \textbf{Split} & \textbf{\# Speakers} &  \textbf{\# Utterances}  & \textbf{\# hrs}  \\
         \midrule
         Test & 199 & 5,699 & 9 \\
         Train & 1,246 & 956,645 & 1158\\
         Val. & 24 & 358 & 0.64 \\
         \bottomrule
    \end{tabular}
    \caption{Final size of our prompted speech dataset.}
    \label{tab:si_asr_size_stats}
\end{table}
\vspace{-0.1cm}

\subsubsection{\textbf{Conversational Speech Test Set}}
\label{sec:realconvo_dataset}
For real world usage, it is also important to evaluate performance of models on unprompted spontaneous speech which is where ASR is most valuable. 
Spontaneous, conversational speech deviates from prompted speech in several key aspects %
e.g., individuals tend to speak at a higher rate with less precise enunciation. Additionally, self-repair of utterances, word repetition, and the use of a broader, context-specific vocabulary are common \cite{quaglio2008conversation}. However, curating datasets of spontaneous speech is time intensive and expensive, as it requires annotators to carefully listen to and transcribe what is being said. This problem is heightened in the case of transcribing disordered speech as it requires a careful ear and expertise (in some cases).  %
Thus, to evaluate on unprompted speech, we compiled a conversational speech test set with the support of a pool of 29 participants with disordered speech. Table~\ref{tab:rc_diversity_stats} reports diversity metrics for this pool. This set is less balanced than the \textit{prompted speech set} due to the smaller participant pool. 
The audio was scrubbed of personally identifiable information (PII) and subsequently transcribed by speech-language pathologists. %
This test set encompasses over 1,500 utterances. More details of the test set can be found in~\cite{tobin2024automatic}.%

\begin{table}[th]
    \centering
    \begin{tabular}{l|l}
         \toprule \midrule
         Category & Percentage (\# speakers)   \\
         \midrule
         \multirow{2}{*}{Gender} & male: 69\% (20) \\
                & female: 31\% (9) \\
         \midrule
         \multirow{3}{*}{Severity} & mild: 24.1\% (7) \\
                  & moderate: 37.9\% (11) \\
                  & severe: 37.9\% (11) \\
        \midrule
        Etiology & ALS: 41.4\% (12) \\
        (top-3)  & Cerebral Palsy: 17.2\% (5) \\
                 & Vocal Cord Paralysis: 6.9\% (2) \\
        \bottomrule
    \end{tabular}
    \caption{Diversity of our \textbf{conversational speech test set}.}
    \label{tab:rc_diversity_stats}
\end{table}

\subsection{Model training and supervised fine-tuning}

The USM ASR systems are first pretrained on large amounts of unpaired data and then fine-tuned on supervised ASR data. In this work we start with a pretrained checkpoint
and then repeat the supervised training step adding the training split of our \textit{prompted speech data} to the paired ASR data. For comparison, the training split of our prompted speech dataset is $\sim$1k hours (Table~\ref{tab:si_asr_size_stats}),
while the other data in the supervised training step is $\sim$100k hours. %

Of the different USM variants, we chose to train the \textit{USM-CTC} model since we are interested in long-form performance on multilingual test sets %
(typical speech) as well as our conversational speech test set (disordered speech). 
During training, batches are sampled from the various paired ASR datasets (described in~\cite{zhang2023google}). We keep the sampling weights of the original paired ASR datasets unchanged and experiment with different sampling weights for newly added training split of our prompted speech dataset and renormalized the weights so they sum to $1.0$. %
Since we're interested in training a generic ASR model, not specifically optimized for any domain, we follow~\cite{zhang2023google} to train the model for a fixed 
300k steps\footnote{In ~\cite{zhang2023google} ASR fine-tuning was run for 100k steps. We trained additional steps to study whether overfitting would happen to the disordered speech data over time. We saw similar results at checkpoint 100k to what is presented at 300k. Namely, around 20\% improvement on both atypical test sets with no significant performance impact on typical speech test sets.} instead of using a development set for finding an ideal checkpoint.

\section{Results}

\subsection{Adding disordered speech data is a win-win}

\begin{figure}
    \centering
    \includegraphics[width=0.7\linewidth]{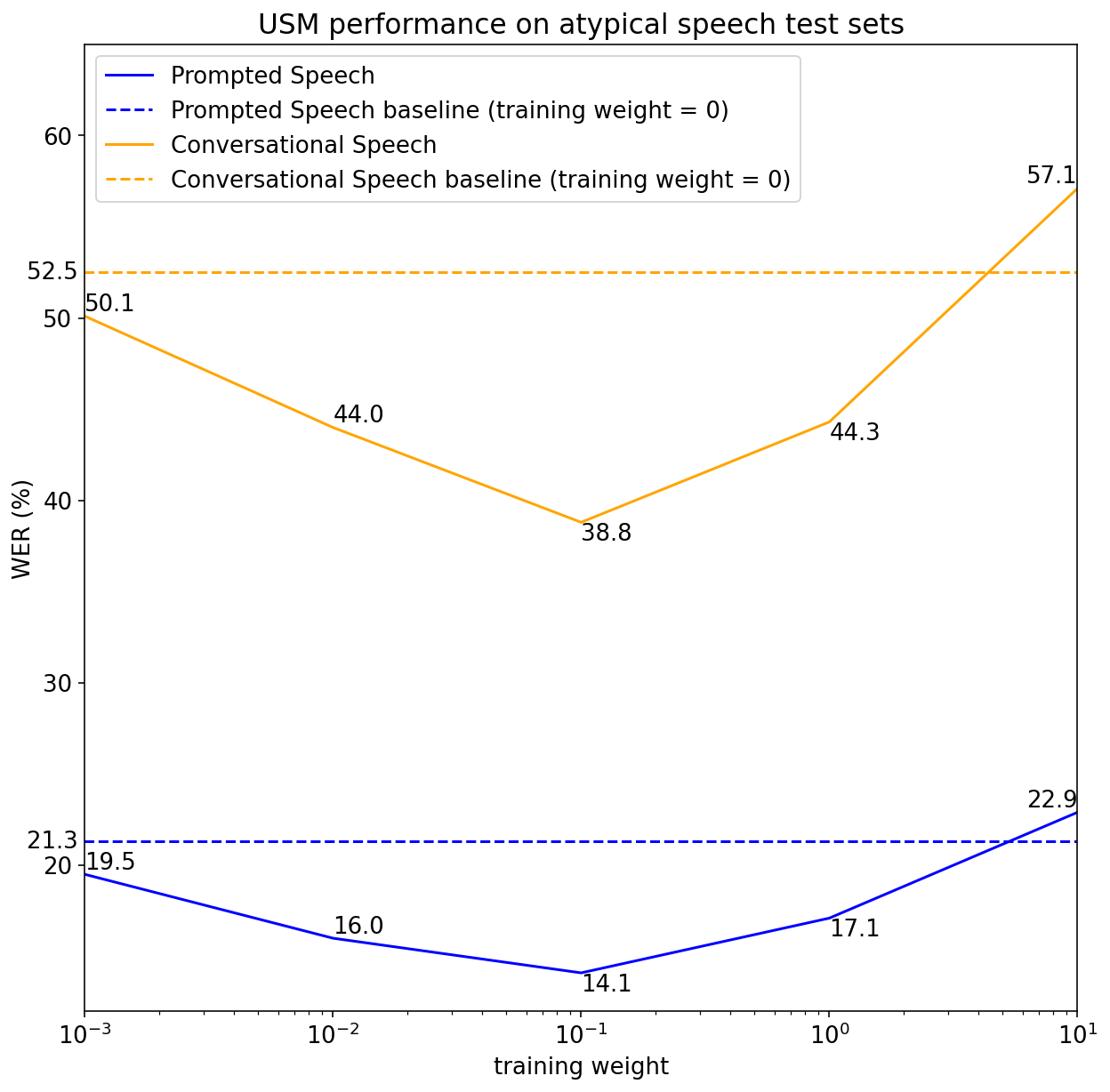}
    \caption{Comparison of performance on test sets, measured by WER on y-axis, based on different training weights.
    }
    \label{fig:weight_comparison}
\end{figure}

Table~\ref{tab:euph_test_set_results_breakout} shows that WER is significantly reduced across all severity levels, amounting to an overall WER reduction of 33\% in the best setting with a mixture weight of $0.1$ on our prompted speech dataset.
The impact of different mixture weights for our datasets can be seen 
in Figure~\ref{fig:weight_comparison}. This figure also compares the addition of our data to the performance of the baseline USM (weight of $0.0$). A weight of $0.1$ consistently gave the best results, across both test sets, and also across all severities reported in the prompted speech dataset (see Table~\ref{tab:euph_test_set_results_breakout}).

\begin{table}[]
    \centering
    \begin{tabular}{l|l|l|l}
         \toprule
         Training data weight & \textbf{Mild}  &  \textbf{Moderate} & \textbf{Severe} \\
         \midrule \midrule
         0 (Baseline) & 10.7 & 29.5 & 48.1\\
         \midrule
         0.001 & 9.7 & 26.5 & 44.5 \\
         0.01 & 8.2 & 21.9 & 35.2 \\
         0.1 & \textbf{7.3} & \textbf{19.6} & \textbf{31.3}\\
         1.0 & 10.3 & 23.1 & 33.8\\
         10.0 & 15.1 & 29.8 & 42.6\\
         \bottomrule
    \end{tabular}
    \caption{Training with a small weight on prompted speech set shows WER improvements on all severity levels. %
    }
    \label{tab:euph_test_set_results_breakout}
\end{table}

Finally, Table~\ref{tab:typical_test_set_results} shows that performance on typical speech, measured on a multilingual test set of English and 18 head languages, %
and Librispeech, is unaffected by the addition of atypical speech data in most cases and only begins to degrade performance at the most extreme weight of $10.0$. Notably, there is a minor improvement on the Librispeech-Other test set at the 0.001 and 0.01 weights, suggesting a positive impact on noisier test sets.

\begin{table}[]
    \centering
    \begin{tabular}{l|l|l|l|l}
         \toprule \midrule
        \multirow{2}{*}{\textbf{Training data weight}} & \multicolumn{2}{c|}{\textbf{Multilingual test}} &  \multicolumn{2}{c}{\textbf{Librispeech}} \\
         &  en-US & 18 langs.  & Clean & Other \\ \midrule
         
         0 (Baseline) & \textbf{13.5}  & \textbf{19.4}  & \textbf{2.4} & 4.6\\
         0.001 & \textbf{13.5} & \textbf{19.4} & \textbf{2.4} & \textbf{4.5}\\
         0.01 & \textbf{13.5} & \textbf{19.4}  & \textbf{2.4} & \textbf{4.5}\\
         0.1 & \textbf{13.5} & \textbf{19.4}  & \textbf{2.4} & 4.6 \\
         1.0 & \textbf{13.5} & \textbf{19.4}  & \textbf{2.4} & 4.6 \\
         10.0 & 13.6 & 19.5 & 2.6 & 4.8 \\
         \hline
    \end{tabular}
    \caption{WER results on typical speech benchmarks do not show degradation. \textbf{bold} shows lowest values. %
    }
    \label{tab:typical_test_set_results}
\end{table}

\subsection{Headroom and comparison to personalization}

Personalized models, fine-tuned for each user, have shown the best performance on atypical speech datasets~\cite{green2021}. These also set the upperbound for SI-ASR models in order to provide users with the best experience while incurring little maintenance costs.  Personalization has usually been done on smaller models e.g., RNN-T~\cite{graves2013speech}. So to see how much progress was made possible by the proposed training recipe, and measure the scope for improvement we compare our USM SI-ASR model against both a personalized RNN-T model and an SI-ASR version trained on the same RNN-T model architecture. Further, comparing the same training recipe on USM vs RNN-T helps elucidate the effect of model size and architecture.
The RNN-T model consists of 
8 LSTM encoder layers and 2 LSTM decoder layers, totalling $\sim$140M parameters. Following the finetuning recipe in~\cite{green2021}, we start from the speaker-independent base model pre-trained on 162k hours of typical speech~\cite{narayanan2019recognizing} and only update the first 5 encoder layers.

The RNN-T SI-ASR model was trained with the same prompted speech training set as the USM experiments, while the personalized models were trained on individual speakers' training, development and test sets, as laid out in \cite{green2021}. %
The data for both the personalization and SI-ASR experiments overlap significantly and come from pre-defined splits from Euphonia~\cite{macdonald2021}. 
For a fair comparison, %
we report results on a subset of utterances of the respective test sets which overlap. This set consists of $1,740$ utterances. Table~\ref{tab:comp_siasr_p13n} compares the results of the models. %

In the case of RNN-T SI-ASR model, while there's 31\% reduction in WER relative to the RNN-T baseline, the personalized model has a much lower WER at 11.3. However, with the USM models, the baseline itself is already comparable to the RNN-T SI-ASR model (28.8 vs 26.4), and tuning with disordered speech data brings down the error further to 17.5, closing 64\% of the gap between the USM baseline (28.8) and the RNN-T personalized model (11.3).
This shows that while the USM model architecture helped reduce WERs from 38.2 to 28.8 on disordered speech, the contribution of the prompted speech training set in the recipe helped bring this error down significantly to 17.5, narrowing the gap between SI-ASR and personalized models. This demonstrates that even 
for individuals where there is no personalization data available (or recording it would be too cumbersome), the USM SI-ASR model can recognize their speech better.

\begin{table}[]
    \centering
    \begin{tabular}{l|l|l}
         \toprule \midrule
         \textbf{Model} & \textbf{WER} & \textbf{WERR} \\
         \midrule
         USM baseline & 28.8 & - \\
         USM SI-ASR & \textbf{17.5} & 39\%  \\
         \midrule
         RNN-T baseline & 38.2 & - \\
         RNN-T SI-ASR & 26.4 & 31\% \\
         RNN-T Personalized & \textbf{11.3} & 71\%\\
         \bottomrule
    \end{tabular}
    \caption{Results comparing unadapted, speaker independent and personalized models, evaluated on a subset of prompted speech test set.}
    \label{tab:comp_siasr_p13n}
\end{table}

\section{Conclusion and Future Directions}

Overall, our work shows how a high-performing ASR model targeted at typical speech can  be trained such that it also performs significantly better on long-tail speech patterns -- disordered speech in our scenario here -- without negatively impacting it's original performance. 
Our experiments on the USM~\cite{zhang2023google} ASR system shows that with relatively small amounts of disordered speech samples, constituting $<1\%$ of the fine-tuning data, when added to the overall supervised fine-tuning data mix can have a large performance improvement on disordered speech without affecting typical speech performance. Further, our experiments comparing the USM SI-ASR model with personalized models indicate that the proposed approach considerably reduces the gap between the baseline model and the personalized model, bringing SI-ASR model performance much closer to  personalized models than possible previously.

This work also studies the effect of adding / weighting the disordered speech data in different proportions. Investigating the ideal training data mixture is an interesting future direction. 
The more important takeaway is that, the  disordered speech and typical speech data were grouped in the same training sets without any special labeling or signal to distinguish the two types of speech.
These findings strongly suggest that incorporating long-tail speech patterns %
can be done relatively easily earlier in the training process. This should be a strong consideration when developing large models to 
advance fairness and accessibility in speech technology, particularly as it aligns with minimal additional resource requirements.
In accessibility, there are typically few one-size-fits-all solutions. Individuals who may have severely impaired speech or require high accuracy for specific domains, such as in medical, technical or business settings, may still be served best by a personalized model. On the other hand, for individuals with milder severity of speech disorders, the type of speaker independent model we presented above, where disordered speech data was added to the training data, may be already good enough for many applications.

In the context of large multi-lingual ASR models (like USM), another interesting future line of research could be to acquire more data in multiple languages to leverage the multilingual ASR capabilities. Specifically, datasets of atypical speech outside of English can be gathered and similarly utilized in training to further improve the models' usefulness for individuals who do not speak English. A large model that performs well on disordered speech may be useful for pseudo-labeling more disordered speech recordings. This data can then be used in noisy student training and further improve a speaker independent model. 

\section{Limitations - Data availability}
The datasets presented in this paper unfortunately cannot be made  available to be shared publicly in adherence to the written consent of Project Euphonia participants, though individual participants can request their own data. %

However, an alternative source of disordered speech data is available to researchers through the Speech Accessibility Project~\cite{SAP} %
a collaboration between researchers at the University of Illinois Urbana-Champaign (UIUC), and five technology companies, including Google. The SAP aims to collect and curate datasets of impaired speech (both prompted and spontaneous) that will be made available to requestors who sign the UIUC's Data Use Agreement and whose application is deemed aligned with the program's objectives by UIUC. SAP's data collection process has been inspired by the data collection procedure of the Euphonia corpus~\cite{macdonald2021} and insights gained by research on it. 

\bibliographystyle{IEEEtran}
\bibliography{mybib}

\end{document}